\documentclass[twocolumn,prb,aps,superscriptaddress,showpacs]{revtex4}
\usepackage{graphicx}
\usepackage{bm}

\begin{document}

\title{Nonlinear optics of semiconductors under an intense terahertz field}
\author{Ren-Bao Liu}
\affiliation{Center for Advanced Study, Tsinghua University,
 Beijing 100084, China}
\author{Bang-Fen Zhu}
\affiliation{Center for Advanced Study, Tsinghua University,
 Beijing 100084, China}

\begin{abstract}
A theory for nonlinear optics of semiconductors in the presence
of an intense terahertz electric field is constructed based on
the double-line Feynman diagrams, in which the nonperturbative
effect of the intense terahertz field is fully taken into account
through using the Floquet states as propagating lines in the
Feynman diagrams.
\end{abstract}

\maketitle

\section{Introduction}

Since early 1990's, thanks mainly to the emergence of free
electron lasers operating in the terahertz (THz)
waveband,\cite{FEL} the interaction between semiconductors and a
strong THz field has been brought under intensive investigations.
Nonlinear transport\cite{iBO,domain,tunneling,self} and linear
optics\cite{DFKE,XDFKE,linear,sideband,sideband1,gap,XS} are the
two main themes of these investigations (here we use the
terminology of ``linear optics'' or ``nonlinear optics'' in the
sense that the intense THz field is treated as a part of the
system but not an optical excitation, otherwise, if the THz field
is viewed as an external optical field, even the so-called
``linear optics'' here would be highly nonlinear). To thoroughly
understand the physics in THz-field-driven semiconductors, as
well as to develop novel devices based on these systems,
nonlinear optical spectroscopies are a powerful and sometimes
necessary method due to their accessibility both in ultrafast
time-resolution and in multi-frequency mixing. For example, the
four-wave mixing spectroscopy has been adopted to study the
effect of the strong THz field induced by Bloch oscillation in
biased semiconductor superlattices,\cite{Lyssenko} and a theory
based on Floquet states\cite{Floquet} of time-periodic systems
has been developed to consider the non-perturbative effects of
the THz dipole field.\cite{Liu} Recently, the
difference-frequency processes were proposed to generate THz
emission, and the estimated strength of THz field could be of the
order of kV/cm,\cite{Kim,Liu1} which is so large that the
feedback effect of the THz field on the nonlinear
difference-frequency process may be important. So, as a common
theoretical basis, a nonlinear optical theory of semiconductors
in the presence of an intense THz field is desired.

To construct such a theory, it is essential to include the
non-perturbative driving of the THz field. To this end, a good
starting point is the eigen states of the THz-driven systems, the
Floquet states,\cite{Floquet} which have the non-perturbative
effect of the THz field fully included. In fact, a compact theory
for linear optics of THz-driven semiconductors has been formulated
in the Floquet-state basis.\cite{linear} In next section, the
general formalism for nonlinear optics of semiconductors under a
strong THz field will be constructed with the double-line Feynman
diagrams frequently used in textbooks.\cite{Shen,Mukamel} In
Section \protect\ref{examples}, some examples will be given to
illustrate how to calculate the nonlinear optical susceptibility
from the Feynman diagrams. And the conclusions are given in the
last section.

\section{General theory}

The system to be considered is a semiconductor irradiated by an
intense cw THz laser. The Hamiltonian of this system under excitation
of additional weak lasers can be expressed as
\begin{equation}
H=H_0(t)-\sum_p \hat{\bm\mu}\cdot{\mathbf F}_p(t),
\end{equation}
where $H_0(t)$ is the unperturbated Hamiltonian of the
semiconductor with the THz-field-driving included, $\hat{\bm\mu}$ is
the dipole operator, and
\begin{equation}
{\mathbf F}_p(t,{\mathbf R})=
{\mathbf F}_pe^{i{\mathbf K}_p\cdot{\mathbf R}-i\Omega_p t}+{\rm c.c.}
\end{equation}
is the pertubative optical
field.

With the density matrix of the system denoted by $\hat{\rho}(t)$,
the optical polarization is ${\mathbf P}(t)={\rm Tr}\left[
\hat{\rho}(t)\hat{\bm \mu}\right]$. As the THz field, with photon
energy much smaller than the band gap, induces no inter-band
excitation, the system is assumed in the semiconductor ground
state before optical excitation, i.e.
$\hat{\rho}(-\infty)=|0\rangle\langle 0|$.  Thus the $j$th
component of the $N$th order [$\chi^{(N)}$] nonlinear optical
response to the optical fields is\cite{Shen,Mukamel}
\begin{widetext}
\begin{eqnarray}
&& {P}_j^{(N)}(t)=\sum_{{\rm P}\{j_1,j_2,\ldots,j_N,j\}}
\int^{+\infty}_{-\infty}
 {\rm Tr}\Big[ U(t,t_{{n}})\theta(t-t_{n}) {i\over \hbar}{\hat{\mu}}_{j_n}F_{j_{n}}(t_{n})
\times U(t_{{n}},t_{{n-1}})\theta(t_{n}-t_{n-1})
{i\over \hbar}{\hat{\mu}}_{j_{n-1}}{F}_{j_{n-1}}(t_{n-1}) \nonumber \\
&&  \phantom{=P\{j_1,j_2,j_N,j\}} \cdots \times
U(t_{{2}},t_{{1}})\theta(t_{2}-t_{1}) {i\over
\hbar}{\hat{\mu}}_{j_1}{F}_{j_{1}}(t_{1}) |0\rangle \langle0|
(-{i\over \hbar}){\hat{\mu}}_{j_{n+1}}{F}_{j_{n+1}}(t_{{n+1}})
U(t_{{n+1}},t_{{n+2}})\theta(t_{n+2}-t_{n+1}) \nonumber \\ &&
\phantom{=P\{j_1,j_2,j_N,j\}}
\cdots
 \times
(-{i\over \hbar}){\hat{\mu}}_{j_{N}}{F}_{j_{N}}(t_{{N}})
U(t_{{N}},t)\theta(t-t_{N})\times  {\hat{\mu}}_j \Big]
dt_1dt_2\cdots dt_N \nonumber \\ && \phantom{P\{j_1,j_2,\}} \equiv
\int^{+\infty}_{-\infty}
\chi^{(N)}_{j;j_1,j_2,\ldots,j_N}(t;t_1,t_2,\ldots,j_N)
F_{j_{1}}(t_{1})F_{j_{2}}(t_{2}) \cdots F_{j_{N}}(t_{N})
dt_1dt_2\cdots dt_N, \label{Pt}
\end{eqnarray}
\end{widetext}
where the summation is over all permutations of the indices as
indicated, and
$$
U(t,t')\equiv\hat{\rm T}e^{-{i\over
\hbar}\int_{t'}^tH_0(t_1)dt_1}=U^{\dag}(t',t)
$$
is the unperturbed propagator of the system. The system in the
presence of an THz field is time-periodic, i.e.
$H_0(t)=H_0(t+T)$, where $T\equiv 2\pi/\omega$ with $\omega$
denoting the angular frequency of the THz field. The eigen states
of the time-periodic Hamiltonian is the Floquet states
$\{|q,t\rangle\}$, which are time-periodic and satisfy the
secular equation
\begin{equation}
\left[H_0(t)-i\hbar\partial_t\right]|q,t\rangle
={}{E}_q|q,t\rangle ={}{E}_q|q,t+T\rangle, \label{qqq}
\end{equation}
where ${}{E}_q$, in analogy of quasi-momentum of Bloch states, is
termed quasi-energy. Obviously, the sidebands of the Floquet
states $|q,m,t\rangle\equiv \exp(im\omega t)|q,t\rangle$ are also
eigen states of Eq. (\protect\ref{qqq}) with quasi-energy
${}{E}_{q,m}\equiv{}{E}_q+m\hbar\omega$. Now the propagator can be
expanded into the Floquet states as
\begin{equation}
U(t,t') =|q,t\rangle e^{-{i\over \hbar}{}{E}_q(t-t')} \langle
q,t'|, \label{UU}
\end{equation}
(hereafter all superscripts and subscripts appearing only on the righthand
side of an equation are assumed dumb indices to be summed over). The dipole
matrix element between the Floquet states is
\begin{equation}
{\bm\mu}_{q;q'}(t)=\langle q,t|{\hat{\bm\mu}}|q',t\rangle
=e^{im\omega t}{\bm\mu}_{q,m;q'}, \label{mu}
\end{equation}
where
$$
{\bm\mu}_{q,m;q'}\equiv T^{-1}\int^T_0 \langle
q,m,t|{\hat{\bm\mu}}|q',t\rangle {dt} = {\bm\mu}_{q;q',-m}
$$
is the time-average of the dipole matrix element. Thus we have
\begin{eqnarray}
U(t',t)\hat{\bm\mu} & = &|q',m',t'\rangle e^{-{i\over
\hbar}{}{E}_{q',m'}(t'-t)}
 {\bm\mu}_{q',m';q}\langle q,t|,
\label{UUmu}  \\
\hat{\bm\mu}U(t,t') & = & |q,t\rangle {\bm\mu}_{q;q',m'}
  e^{-{i\over \hbar}{}{E}_{q',m'}(t-t')}
 \langle q',m',t'|. \label{muUU}
\end{eqnarray}

\begin{figure}[t]
\includegraphics[height=12cm,width=7.5cm, bb=50 50 520 750,
 clip=true]{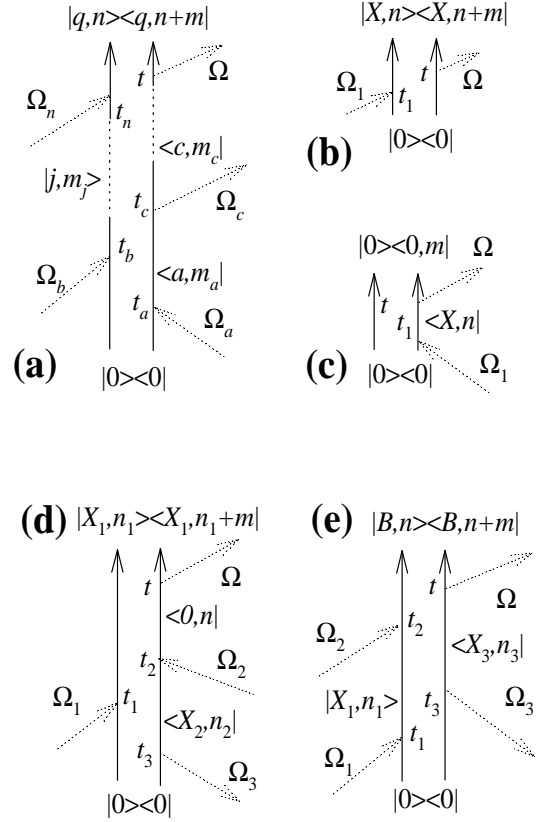}
\caption{(a) The general form of double-line Feynman diagrams for
nonlinear optics of semicondcutors in the presence of an intense
THz electric field. (b), (c) The double-line Feynman diagrams
for linear optics of THz-field-driven semicondcutors. (d),(e)
Two of the double-line Feynman diagrams for $\chi^{(3)}$ four-wave
mixing in THz-field-driven semiconductors.}
\label{fig1}
\end{figure}

With Eq. ({\protect\ref{UUmu}) and ({\protect\ref{muUU}) used
respectively on the left and right sides of $|0\rangle\langle 0|$
in Eq. (\protect\ref{Pt}), the nonlinear optical response can be
derived. The double-line Feynman diagrams can be used to assist
such a derivation as well as in static cases. For this particular
time-periodic system, the rules of the Feynman diagrams for the
optical response $P_j^{(N)}(t)$ are [see Fig. \protect\ref{fig1}
(a)]:
\begin{enumerate}
\item The evolution of the diagram starts from the ground state
 $|0\rangle\langle 0|$ at $t=-\infty$ and ends at $t$ with the final
 Floquet-state density matrix $|q,n\rangle \langle q,n+m|$.
\item The $N$ interaction vertices at $t_p<t$ ($p=1,2,\ldots,N)$
 consist of a photon line (the dotted arrow) with frequency
 $\Omega_p$, the Floquet state before interaction ($|a,m_a\rangle$ on
 the left branch or $\langle a,m_a|$ on the right branch), and the
 state after interaction ($|c,m_c\rangle$ on the left branch or
 $\langle c,m_c|$ on the right branch). The photon lines pointing to
 inside and outside represent photon absorption and emission processes, respectively.
 The $p$th vertex contributes a dipole matrix element
 ${i\over \hbar}{\bm \mu}_{c,m_c;a,m_a}$
 (or $-{i\over \hbar}{\bm \mu}_{a,m_a;c,m_c}$) if it's on the left (or right) branch, and
 a factor from the optical field
 ${\mathbf F}_p\exp(-i\Omega_p t_p+i{\mathbf K}_p\cdot{\mathbf R})$
 for photon absorption or
 ${\mathbf F}_p^*\exp(i\Omega_p t_p-i{\mathbf K}_p\cdot{\mathbf R})$
 for photon emission.
\item The photon emission vertex at final time $t$ contributes the factor
 $(\mu_j)_{c,m_c;q,n+m}e^{-im\omega t}$ where $\langle c,m_c|$ is the
  state before photon emission.
\item The double-line between two neighbor vertices at $t_1$ and $t_2$
 ($t_1<t_2$) represents the unperturbed propagator of the density matrix.
 If the state associated with the double-line is $|b,m_b\rangle\langle c,m_c|$,
 the factor due to the propagator is
 $$
 \theta(t_2-t_1)
 e^{-{i\over \hbar}\left({}{E}_{b,m_b;c,m_c}-i\Gamma_{b,m_b;c,m_c}\right)(t_2-t_1)},
 $$
 where ${}{E}_{b,m_b;c,m_c}\equiv {}{E}_{b,m_b}-{}{E}_{c,m_c}$ is the transition
 energy between the Floquet states, and $\Gamma_{b,m_b;c,m_c}/\hbar$ is the
 relaxation rate of the density matrix due to the interaction with environment.
\item For each diagram, all factors described above should be
multiplied together and all Floquet states and their sidebands
are summed over. Then all different diagrams, which are
determined both by how the $N$ vertices are grouped into two
branches and by the time-ordering, should be summed. For a
certain optical configuration (which determines each vertex is a
photon absorption or emission process), there are totally $2^N
N!$ different diagrams, among which there are $2^N$ elementary
diagrams and the others can be derived from them by permutation
of the $N$ vertices. And still, there are $2^N$ different optical
configurations. In a certain experiment, only one optical
configuration needs to be considered.
\end{enumerate}

The nonlinear optical susceptibility in frequency domain is defined as
\begin{widetext}
\begin{eqnarray}
\chi^{(N)}_{j;j_1,j_2,\ldots,j_N}
(\Omega;\Omega_1,\Omega_2,\ldots,\Omega_N) = \int  dt_1dt_2\cdots
dt_N \chi^{(N)}_{j;j_1,j_2,\ldots,j_N}(t;t_1,t_2,\ldots,j_N)
e^{i\Omega t-i\Omega_1t_1-i\Omega_2t_2\cdots-i\Omega_Nt_N}.
\end{eqnarray}
\end{widetext}
The rules for constructing susceptibility in frequency domain from the Feynman
diagrams can be easily derived from the rules above:
\begin{enumerate}
\item The interaction vertex, consitituted by the photon line with
 frequency $\Omega_p$, the initial state $|a,m_a\rangle$
 (or $\langle a,m_a|$), and the final state $|c,m_c\rangle$
 (or $\langle c,m_c|$), contributes the factor
 ${i\over \hbar}(\mu_{j_p})_{c,m_c;a,m_a}$
 (or $-{i\over \hbar}(\mu_{j_p})_{a,m_a;c,m_c}$)
 if the vertex is on the left (or right) branch .
\item The factor associated with the double-line state
 $|b,m_b\rangle\langle c,m_c|$ between $t_1$ and $t_2$ ($t_1<t_2$) is
 $$
 -i\hbar\Big({}{E}_{b,m_b;c,m_c}-i\Gamma_{b,m_b;c,m_c}
 -\sum_{t_n<t_2}\Omega_n\Big)^{-1}.
 $$
\item The factor associated with the final photon emission vertex
at $t$ is
 $$
 2\pi\delta\Big(\Omega-m\omega-\sum_{n=1}^N\Omega_n\Big)
 (\mu_j)_{c,m_c;q,n+m},
 $$
 where $\langle c,m_c|$ is the state before photon emission.
\item All factors are multiplied together and all intermediate
 Floquet states and their sidebands are summed over. Then all
 possible diagrams are summed. For a certain optical configuration,
 the frequency $\Omega_p$ is positive or negative depending on whether
 the corresponding vertex is an absorption or emission process.
\end{enumerate}

From the rules described above, we can see that the nonlinear
optical response of THz-field-driven semiconductors, or generally
speaking, time-periodic systems, takes the form very similar to
the textbook formalism for static systems.\cite{Shen,Mukamel} The
difference lies on three aspects: First, the dipole matrix
element here is the time-average of that between Floquet states.
Secondly, the THz field induces new resonances at
THz-photon-assisted transitions, as the sidebands of the Floquet
states act as intermediate states in the nonlinear optical
process . And thirdly, there is an extra dynamic phase factor
$e^{-im\omega t}$ at the final photon-emission vertex, which
makes the energy of the emitted photon differ from the total
input energy by an integer multiple of the THz-photon energy,
corresponding to the physical process of THz-photon-sideband
generation.\cite{sideband,sideband1} In the time-periodic
systems, the energy-conserving condition is relaxed to the
quasi-energy conservation. We reiterate that the nonperturbative
effect of the THz field has been fully included through the
renormalization of dipole matrix element and transition energy.

\section{Examples} \label{examples}

As an illustrative example, the Feynman diagrams for linear optics are
plotted in Fig. \protect\ref{fig1} (b) and (c). From the rules for the
Feynman diagrams, the linear susceptibility can be formulated as
\begin{widetext}
\begin{eqnarray}
\chi^{(1)}_{j;j_1}(\Omega;\Omega_1)=
2\pi\delta(\Omega-\Omega_1-m\omega) \left[
  {\left(\mu_j\right)_{0;X,n+m} \left(\mu_{j_1}\right)_{X,n;0}
  \over {}{E}_{X,n;0}-i\Gamma_{X,n;0}-\Omega_1 }
- { \left(\mu_j\right)_{X,n;0,m} \left(\mu_{j_1}\right)_{0;X,n}
   \over {}{E}_{0;X,n}-i\Gamma_{0;X,n}-\Omega_1}
\right],
\end{eqnarray}
\end{widetext}
where $|X,n\rangle$ denotes the THz-photon sidebands of the
Floquet-state excitons and $|0,n\rangle$ denotes the sidebands of
the ground state . This result is identical to that in Ref.
\onlinecite{linear} derived with non-equilibrium Green's function
technique.

Now we consider another example, the $\chi^{(3)}$ four-wave
mixing, in which three input beams propagate in the directions
${\mathbf K}_1$, ${\mathbf K}_2$, and ${\mathbf K}_3$,
respectively, and the signal is detected in the direction
${\mathbf K}_1+{\mathbf K}_2 -{\mathbf K}_3$. Corresponding to
this optical configuration, there are 48 different Feynman
diagrams, among which only 16 are under resonant excitation
condition. We calculate two typical diagrams as examples of
resonant excitation of excitons and bi-excitons (exciton
molecules constituted by two excitons). Fig. \protect\ref{fig1}
(d) is a Feynman diagram for resonant excitation of excitons,
which contributes to the susceptibility as
\begin{widetext}
\begin{eqnarray}
{2\pi\delta(\Omega-\Omega_1-\Omega_2+\Omega_3-m\omega)
 \left(\mu_{j_3}\right)_{0;X_2,n_2}
 \left(\mu_{j_2}\right)_{X_2,n_2;0,n}
 \left(\mu_{j}\right)_{0,n;X_1,n_1+m}
 \left(\mu_{j_1}\right)_{X_1,n_1;0}
  \over
\left({}{E}_{X_1,n_1;0,n}-i\Gamma_{X_1,n_1;0,n}-\Omega_1-\Omega_2+\Omega_3\right)
\left({}{E}_{X_1,n_1;X_2,n_2}-i\Gamma_{X_1,n_1;X_2,n_2}-\Omega_1+\Omega_3\right)
\left({}{E}_{0;X_2,n_2}-i\Gamma_{0;X_2,n_2}+\Omega_3\right)} ,
\nonumber
\end{eqnarray}
and Fig. \protect\ref{fig1} (e) is a diagram for resonant
excitation of bi-excitons, which contributes to the
susceptibility as
\begin{eqnarray}
{-2\pi\delta(\Omega-\Omega_1-\Omega_2+\Omega_3-m\omega)
 \left(\mu_{j_3}\right)_{0;X_3,n_3}
 \left(\mu_{j}\right)_{X_3,n_3;B,n+m}
 \left(\mu_{j_2}\right)_{B,n;X_1,n_1}
 \left(\mu_{j_1}\right)_{X_1,n_1;0}
  \over
\left({}{E}_{B,n;X_3,n_3}-i\Gamma_{B,n;X_3,n_3}-\Omega_1-\Omega_2+\Omega_3\right)
\left({}{E}_{X_1,n_1;X_3,n_3}-i\Gamma_{X_1,n_1;X_3,n_3}-\Omega_1+\Omega_3\right)
\left({}{E}_{X_1,n_1;0}-i\Gamma_{X_1,n_1;0}-\Omega_1\right)},
\nonumber
\end{eqnarray}
\end{widetext}
where $|B,n\rangle$ denotes the sidebands of Floquet-state
bi-excitons. From the two terms above, we can easily identify the
resonances associated with the THz-photon-assisted Floquet-state
exciton and bi-exciton transitions. The sideband generation, as
indicated by the $\delta$-function with argument containing an
integer multiple of THz-photon energy, accounts for the four-wave
mixing signal out of the excitation spectrum observed in the
numerical calculations in Ref. \onlinecite{Liu}.

\section{Summary}

In summary, based on the double-line Feynman diagrams similar to
the static case in textbooks, we have construct a general theory
for nonlinear optics of semiconductors under an intense THz
field. The basis of the Feynman diagram is the eigen states of
the time-periodic systems, i.e., the Floquet states, so the
non-perturbative effect of the THz field has been fully taken
into account. Many phenomena, including the THz-photon-assisted
exciton or bi-exciton resonances and the THz-photon sideband
generation, are naturally accounted for in this theory.

\acknowledgments{
This work was supported by the National Science Foundation of China.
}

\end{document}